# Instantaneous Measurement of Velocity Fields in Developed Thermal Turbulence in Mercury

Takashi Mashiko,[1] Yoshiyuki Tsuji,[2] Takatoshi Mizuno,[2] and Masaki Sano[1, *]

[1]*Department of Physics, Graduate School of Science, University of Tokyo, Tokyo 113-0033, Japan*
[2]*Department of Energy Engineering and Science, Graduate School of Engineering, Nagoya University, Nagoya 464-8603, Japan*
(Dated: October 23, 2003)

Using ultrasonic velocimetry we measured the vertical profile of the velocity fluctuation in high Rayleigh number thermal convection in a cell with aspect ratio of 0.5, filled with a low Prandtl number fluid, mercury. Intriguing fluctuating dynamics of the mean flow and universal nature of the kinetic energy cascade are elucidated utilizing spectral decomposition and reconstruction. Scaling properties of the structure functions and the energy spectrum are directly calculated without the use of Taylor's frozen-flow hypothesis for the first time. Despite the complex nature of the mean flow, it is found that the energy cascade process exhibits universal laws in thermal turbulence.

Recently, turbulent thermal convection has attracted considerable attention not only because it is ubiquitous in nature and technology but also because it continually offers fundamental questions in the study of turbulence. Among the investigations of high Rayleigh number ($Ra$) thermal turbulence, efforts are made to elucidate the scaling laws for the heat transport [1, 2], its dependence on aspect ratio ($\Gamma$) [3] or on Prandtl number ($Pr$) [4–6], and the scaling laws for the power spectrum [7–10]. Furthermore, following the experimental works which challenged to raise $Ra$ [11–14], the existence of the ultimate state at very high $Ra$ is argued [4, 15–17].

Now it is believed that the global heat transport is affected by the large-scale mean flow, which is found to appear in well developed thermal turbulence [18–23] and to modify the stability of the thermal boundary layers by its shear [23–27]. It is reported that in the highest $Ra$ ($\sim 10^{17}$) experiment in a $\Gamma = 0.5$ cell filled with cryogenic helium gas, the large-scale circulation is no longer unidirectional, but changes its sign frequently [13]. Therefore understanding the structure and the dynamics of the mean flow is of great importance. However, the information on the velocity fields is still scarce especially when both $Ra$ and Reynolds number ($Re$) are high.

Another important issue is to verify the scaling behavior of the fluctuations in thermal turbulence. It is predicted that the scaling exponents for the structure functions are different from those of homogeneous isotropic turbulence [28–30]. Denoting the structure functions of the velocity $v$ and the temperature $T$ by

$$S_n(r) \equiv \langle |\delta v(r)|^n \rangle \equiv \langle |v(x+r) - v(x)|^n \rangle \propto r^{\zeta_n}, \quad (1)$$
$$Q_n(r) \equiv \langle |\delta T(r)|^n \rangle \equiv \langle |T(x+r) - T(x)|^n \rangle \propto r^{\eta_n}, \quad (2)$$

$\zeta_2 = 6/5$ and $\eta_2 = 2/5$ are predicted for thermal turbulence, which are distinct from the values of $\zeta_2 = \eta_2 = 2/3$ for isotropic turbulence. Equivalently for the energy spectrum, $E(k) \propto k^{-11/5}$ is predicted unlike Kolmogorov (K41) spectrum, $k^{-5/3}$, for the isotropic case. However, there are no experimental works which measured the velocity or the temperature at a number of points simultaneously to test the scaling properties in thermal turbulence. The frequency power spectrum $P(f)$, calculated from the local velocity time series, is regarded as equivalent with $E(k)$ by employing Taylor's frozen-flow hypothesis. However, the conditions for Taylor's hypothesis become questionable in a confined geometry and especially when the mean flow is not dominating. Therefore, the instantaneous measurement of the velocity fields is highly desired for the verification of the universal scaling nature in thermal turbulence.

In this letter, we report the instantaneous measurement of the velocity fields in an opaque liquid, mercury, by using ultrasonic velocimetry (UV). We present the first evidence that the velocity structure functions and the energy spectrum have the universal scaling exponents which are distinct from those of isotropic turbulence by direct calculations. We also present intriguing dynamical properties of the large-scale circulations.

We use liquid mercury, because it has low kinematic viscosity ($\nu$) and high thermal diffusivity ($\kappa$), and consequently low $Pr$ ($\sim 0.024$ at 20°C), thus it is an ideal fluid for the study of the ultimate state in which sheared viscous boundary layer becomes thinner than the thermal boundary layer [17, 25]. $Re$ of the mean circulation defined by $Re = \langle v \rangle L/\nu$, with $\langle v \rangle$ being the mean velocity and $L$ the cell size, can reach as high ($Re > 10^5$) as that in helium gas experiment which attained the highest $Ra$.

Now we outline the principle of the UV method, which is detailed by Takeda [31]. A transducer emits ultrasonic pulses (bursts) of several wavelengths repeatedly at an interval of $T$, and receives the echoes scattered by the tracers in the fluid. The echo signal for the $n$-th burst is well approximated by $s_n(t) \simeq A_n(t) \cos[2\pi f_0 t + \phi_n(t)]$ where $f_0$ is the pulse frequency, $t$ is the elapsed time measured after the burst emission, and $\phi_n(t)$ is the phase shift due to Doppler effect. $\phi_n(t)$ can be calculated by complex demodulation. Since the Doppler-shift frequency of the signal $s_n(t)$ is calculated by $\Delta f = \Delta \phi_n(t)/2\pi T = (\phi_{n+1}(t) - \phi_n(t))/2\pi T$, the velocity component of the direction of the acoustic emission at a distance of $x$ is given by $v_n(t) = -c\Delta\phi_n(t)/4\pi f_0 T$, with $t = 2x/c$.

Our experimental setup was detailed elsewhere [12]. The apparatus is a cylindrical cell with chromium plated copper top and bottom plates of $d = 306$ mm in diameter and stainless steel sidewall of $L = 612$ mm in height. The aspect ratio is thus $\Gamma = 0.5$. The temperature drops in the copper plates are subtracted in order to calculate $Ra$ and Nusselt number ($Nu$) [23]. The UV transducer, whose size is 8 mm in diameter (the bare size is 5 mm), is embedded in the top copper plate so that its cylindrical axis coincides with that of the cell ($z$-axis, whose origin is set at the cell center). We mixed the powder of Gold-Palladium alloy into the mercury as tracers. The composition of the alloy was designed so that its density coincides with that of the mercury, 13.48 g/cm$^3$, at 45°C which is the mean temperature of the two plates. The tracers enable us to measure the whole profile of the velocity field along $z$-axis [32]. The ultrasonic velocimeter (UVP-X-1, MET-FLOW) emits pulses of $f_0 = 4$ MHz. In the typical run, 15360 velocity profiles, each of which consists of 128 data points located every $\Delta z$ mm, are recorded every 131 ms [33]. We can vary the longitudinal spatial resolution $\Delta z$ from 0.72 ($\sim 2 \times$ wavelength) to 5.07 mm. The transversal resolution is 5 mm. The resolution of the measurable velocity is 0.692 mm/s.

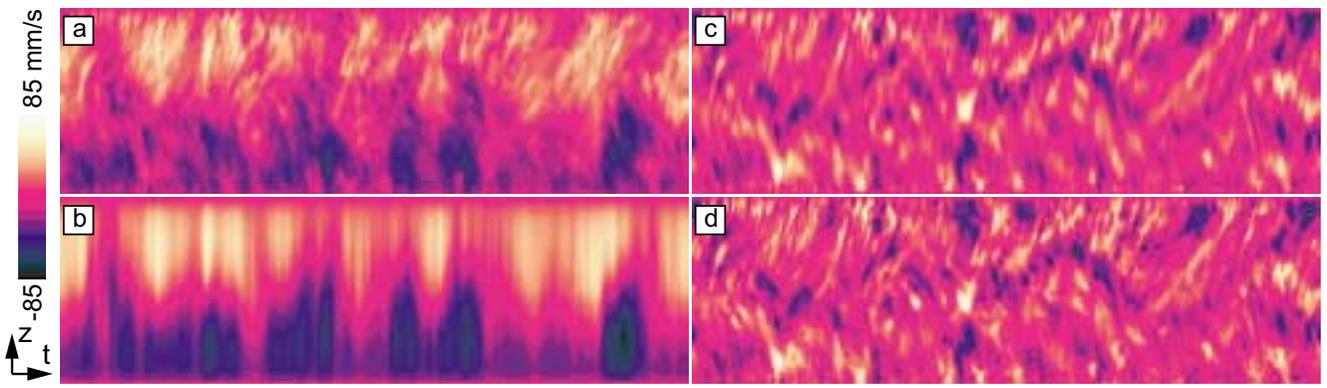

FIG. 1: (color) Spatiotemporal plots of the velocity field at $Ra = 4.79 \times 10^{10}$. The yellow (bright) and the purple (dark) regions correspond to the upward and the downward flows respectively. (a) Original velocity field. (b) Reconstructed field from the mean velocity and the first two eigenmodes. (c) Reconstructed field from the 3rd to the 30th eigenmodes in the scaling range. (d) Residual field, that is, (b) is subtracted from (a). The color bar is for (a), and arbitrary units are used for (b) to (d).

We show a spatiotemporal plot of the velocity field in Fig. 1(a). This plot was obtained at $Ra = 4.79 \times 10^{10}$ and $Re = 1.3 \times 10^5$. In Fig. 1, the horizontal axis denotes the time which spans a period of 134 seconds, while the vertical one is $z$-axis. The yellow (bright) and the purple (dark) regions correspond to the upward and the downward flows respectively. Apparently the upward flow is dominant in the upper-half region while the downward flow is dominant in the lower-half region of the cell.

To characterize the fluctuating nature of the velocity field, we adopt Karhunen-Loève transformation and separate the slow dynamics of the mean flow and the fast cascade dynamics in the scaling range. A successive 15360 velocity profiles with $\Delta z = 5.07$ mm are decomposed into 128 eigenvectors as $v(z, t_n) - \langle v(z) \rangle = \sum_{i=1}^{M} a_i(t_n)\phi_i(z)$ ($n = 1, 2, \ldots, N; N = 15360$), where $\phi_i$'s are the eigenvectors representing the dominant modes in a descending order of $i$ ($i = 1, 2, \ldots, M; M = 128$), and $a_i$'s are the coefficients of expansion. The time-averaged

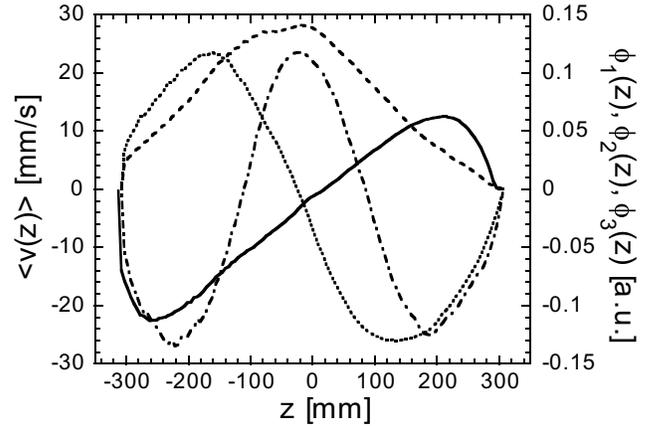

FIG. 2: Time-averaged profile of the velocity, $\langle v(z) \rangle$, and first three eigenvectors $\phi_i(z)$'s. $Ra$ is same as in Fig. 1. The solid curve is $\langle v(z) \rangle$. The dashed, the dotted, and the dot-dashed curves correspond to $\phi_1(z)$, $\phi_2(z)$, and $\phi_3(z)$, respectively.

velocity profile and the first several eigenvectors are shown in Fig. 2.

Figure 1(b) shows the slow dynamics reconstructed by the superposition of the mean profile and the first two eigenmodes. We note that there is no clear periodicity in the dynamics. Figure 1(c) shows the dynamics reconstructed from the 3rd to the 30th eigenmodes which are well in the scaling range shown in Fig. 3. Therefore it visualizes the cascade dynamics in the buoyancy dominating inertial range. Figure 1(d) is the residual dynamics reconstructed by subtracting the lowest modes of Fig. 1(b) from the raw data of Fig. 1(a). As we recognize, the mean profile in Fig. 2 has some asymmetry. Nevertheless, the symmetry is recovered in the scaling range as shown in Fig. 1(c), and the eigenvectors in Fig. 2 are also symmetric. This implies that the scaling range in which a self-similarity holds universally exists in thermal turbulence

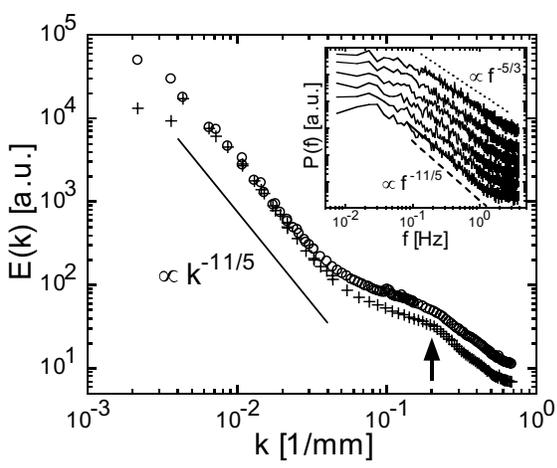

FIG. 3: Energy spectral density $E(k)$ calculated by subtracting the mean profile (circles) and the first two eigenmodes in addition (crosses). By fitting, $\beta = -2.15 \pm 0.02$ and $-2.22 \pm 0.02$ were obtained for $E(k) \propto k^{-\beta}$, respectively. The arrow indicates the lateral resolution. Inset: Frequency power spectra $P(f)$'s at $z = 0, 50, \ldots, 250$ mm (from bottom to top).

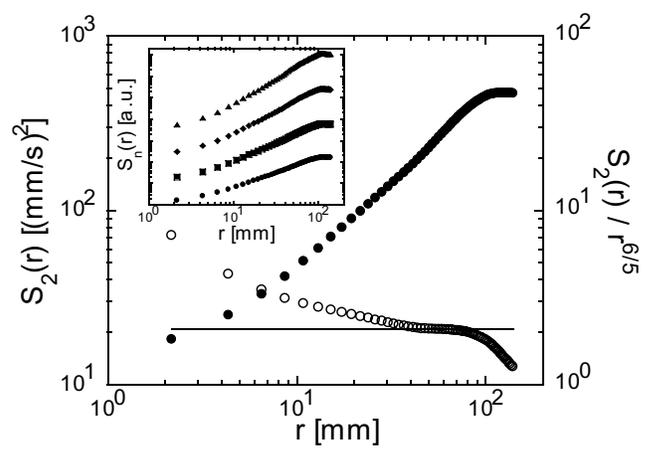

FIG. 4: Second-order structure function $S_2(r)$ (filled circles) and $S_2(r)/r^{6/5}$ (open circles) calculated from the velocity profiles with $\Delta z = 2.17$ mm around the cell center. Inset: Higher order functions, $S_3(r)$ to $S_6(r)$ (from bottom to top).

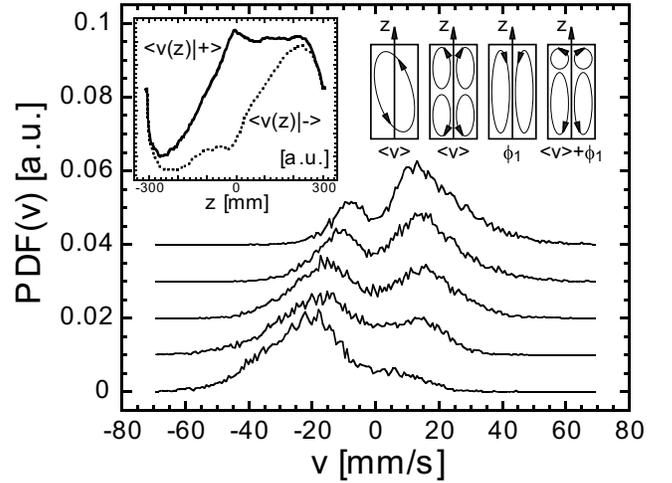

FIG. 5: Probability distribution functions of the velocity at $z = -200, -100, \ldots, 200$ mm (from bottom to top). Inset: (left) Conditional average of the velocity profile, $\langle v(z)|\pm\rangle$ (see text). (right) Imaginary structures of the macroscopic flow.

irrespective of the large-scale dynamics.

Now we show in Fig. 3 the energy spectral density $E(k)$ directly calculated from the results of the multipoint velocity measurements. Measurements were performed over three different $z$-ranges of $127 \times \Delta z$ with $\Delta z = 3.62$, 2.17, and 0.72 mm. In each measurement, the center of the range is set at $z = 0$ mm, and 10240 profiles were sampled. Data from three $\Delta z$'s overlap each other in Fig. 3. The arrow indicates the transversal resolution. To avoid a bias from the mean profile $\langle v(z) \rangle$, we subtracted it from each measurement and calculated $E(k)$ (circles). To be more careful we subtracted the slow dynamics due to the first two eigenmodes together with the mean profile (crosses). For both plots, there exists a scaling range in which the power law $E(k) \propto k^\beta$ holds. By fitting, $\beta = -2.15 \pm 0.02$ and $-2.22 \pm 0.02$ were obtained for the circles and the crosses, respectively. Both are close to $-11/5$ derived in Bolgiano-Obukhov theory [28, 29] for thermal turbulence (represented by the solid line in Fig. 3), which is distinct from $-5/3$ for the isotropic case.

Historically the frequency power spectra $P(f)$'s have been calculated from local time series [9, 10]. In the inset of Fig. 3 we show $P(f)$'s at positions located every 50 mm along $z$-axis, $z = 0, 50, \ldots, 250$ mm, together with the power laws with exponents of $-11/5$ (the dashed line) and $-5/3$ (the dotted line) for comparison. Each $P(f)$ was calculated from the velocity time series consisting of 10240 samples. The slope of $P(f)$ changes with the position. At $z = 0$ mm, the slope is $-2.14 \pm 0.03$ which is close to $-11/5$. But at the cell center $P(f)$ has no relation to $E(k)$ since the mean velocity $\langle v(0) \rangle$ is zero. At $z = 250$ mm, the slope is $-1.72 \pm 0.02$. Here the fluctuation-to-mean ratio, $v_{\text{rms}}/\langle v \rangle$, remains near 1, thus Taylor's hypothesis is not valid in turn. Thus $P(f)$ is different from $E(k)$ in thermal turbulence.

To confirm the scaling properties more firmly, we show in Fig. 4 the second-order structure function $S_2(r)$ and the compensated plot $S_2(r)/r^{6/5}$ as functions of $r$. The structure function was calculated from series of the velocity profiles with $\Delta z = 2.17$ mm around the cell center. The mean profile and the lowest modes were subtracted from the raw data as in Fig. 1(c). We note that there exists a scaling range in which the predicted scaling exponent $\zeta_2 = 6/5$ is observed. This exponent is distinct from that of isotropic turbulence, $\zeta_2 = 2/3$. In the inset we show higher order $S_n(r)$'s. By fitting we obtain $\zeta_2 = 1.18$, $\zeta_3 = 1.58$, $\zeta_4 = 2.03$, $\zeta_5 = 2.29$, and $\zeta_6 = 2.55$.

Finally, we discuss the probability distribution functions

(PDFs) of the velocity. Looking at Fig. 1 carefully, we notice that the velocity at each point changes its sign occasionally. To see this clearly, we show in Fig. 5 the velocity PDFs at positions located every 100 mm along $z$-axis, $z = -200, -100, \ldots, 200$ mm, each of which resulted from 15360 samples. It is seen that the PDFs are characterized by the double-peaked form. Niemela *et al.* obtained a similar PDF by indirect method [13], but its mechanism remains unknown. To elucidate the mechanism, we calculated the conditional average of the velocity profile, $\langle v(z)|\pm\rangle \equiv \langle v(z)|v(0) = v_{\text{peak}}^{\pm}\rangle$, as shown in the left inset of Fig. 5, where $v_{\text{peak}}^{\pm}$ denotes the value at which the PDF takes the peak in the positive or the negative side. We notice that the double-peaked PDFs are originated from the sloshing motion of the mean flow. This slow dynamics of the mean flow can be explained by the interaction among eddies of different scales. From the mean profile $\langle v(z)\rangle$, we can imagine some structures of the macroscopic flow. Two simple examples are shown in the right inset of Fig. 5 (labeled as $\langle v \rangle$). Since 63% of the total fluctuation is contributed by $\phi_1$ and $\phi_2$, the large-scale dynamics is well described by the superposition of $\langle v \rangle$, $\phi_1$, and $\phi_2$. Typical structures described by $\phi_1$ and $\langle v \rangle + \phi_1$ are shown in the inset schematically. As the coefficient $a_1(t)$ changes its sign, the rolls make a sloshing motion which again explains the double-peaked form of the PDFs. Therefore we speculate that the large-scale circulation is just the largest eddy in the cascade, and which has a highly fluctuating nature.

To conclude, the usefulness of the UV method for the measurement of the velocity fluctuation in thermal turbulence has been proved. We expect that this work will activate further theoretical and experimental studies in thermal turbulence.

We acknowledge Y. Takeda for valuable comments and discussions. This work was supported by a Japanese Grant-in-Aid for Scientific Research from the Ministry of Education, Culture, Sports, Science, and Technology (No. 14340119).